\PassOptionsToPackage{unicode}{hyperref}
\PassOptionsToPackage{hyphens}{url}
\PassOptionsToPackage{table,dvipsnames,svgnames,x11names}{xcolor}
\documentclass[12pt]{article}

\usepackage{amsmath,amssymb}
\usepackage{iftex}
\ifPDFTeX
  \usepackage[T1]{fontenc}
  \usepackage[utf8]{inputenc}
  \usepackage{textcomp} 
\else 
  \usepackage{unicode-math}
  \defaultfontfeatures{Scale=MatchLowercase}
  \defaultfontfeatures[\rmfamily]{Ligatures=TeX,Scale=1}
\fi
\usepackage{lmodern}
\ifPDFTeX\else  
\fi
\IfFileExists{upquote.sty}{\usepackage{upquote}}{}
\IfFileExists{microtype.sty}{
  \usepackage[]{microtype}
  \UseMicrotypeSet[protrusion]{basicmath} 
}{}
\makeatletter
\@ifundefined{KOMAClassName}{
  \IfFileExists{parskip.sty}{%
    \usepackage{parskip}
  }{
    \setlength{\parindent}{0pt}
    \setlength{\parskip}{6pt plus 2pt minus 1pt}}
}{
  \KOMAoptions{parskip=half}}
\makeatother
\usepackage{xcolor}
\usepackage{bm}
\usepackage{bbm}
\usepackage{mathrsfs} 
\usepackage{threeparttable}

\makeatletter
\ifx\paragraph\undefined\else
  \let\oldparagraph\paragraph
  \renewcommand{\paragraph}{
    \@ifstar
      \xxxParagraphStar
      \xxxParagraphNoStar
  }
  \newcommand{\xxxParagraphStar}[1]{\oldparagraph*{#1}\mbox{}}
  \newcommand{\xxxParagraphNoStar}[1]{\oldparagraph{#1}\mbox{}}
\fi
\ifx\subparagraph\undefined\else
  \let\oldsubparagraph\subparagraph
  \renewcommand{\subparagraph}{
    \@ifstar
      \xxxSubParagraphStar
      \xxxSubParagraphNoStar
  }
  \newcommand{\xxxSubParagraphStar}[1]{\oldsubparagraph*{#1}\mbox{}}
  \newcommand{\xxxSubParagraphNoStar}[1]{\oldsubparagraph{#1}\mbox{}}
\fi
\makeatother

\usepackage{longtable,booktabs,array}
\usepackage{calc} 
\usepackage{etoolbox}
\makeatletter
\patchcmd\longtable{\par}{\if@noskipsec\mbox{}\fi\par}{}{}
\makeatother
\IfFileExists{footnotehyper.sty}{\usepackage{footnotehyper}}{\usepackage{footnote}}
\makesavenoteenv{longtable}
\usepackage{graphicx}
\makeatletter
\def\maxwidth{\ifdim\Gin@nat@width>\linewidth\linewidth\else\Gin@nat@width\fi}
\def\maxheight{\ifdim\Gin@nat@height>\textheight\textheight\else\Gin@nat@height\fi}
\makeatother
\setkeys{Gin}{width=\maxwidth,height=\maxheight,keepaspectratio}
\makeatletter
\def\fps@figure{htbp}
\makeatother

\makeatletter
\@ifpackageloaded{caption}{}{\usepackage{caption}}
\AtBeginDocument{%
\ifdefined\contentsname
  \renewcommand*\contentsname{Table of contents}
\else
  \newcommand\contentsname{Table of contents}
\fi
\ifdefined\listfigurename
  \renewcommand*\listfigurename{List of Figures}
\else
  \newcommand\listfigurename{List of Figures}
\fi
\ifdefined\listtablename
  \renewcommand*\listtablename{List of Tables}
\else
  \newcommand\listtablename{List of Tables}
\fi
\ifdefined\figurename
  \renewcommand*\figurename{Figure}
\else
  \newcommand\figurename{Figure}
\fi
\ifdefined\tablename
  \renewcommand*\tablename{Table}
\else
  \newcommand\tablename{Table}
\fi
}
\@ifpackageloaded{float}{}{\usepackage{float}}
\floatstyle{ruled}
\@ifundefined{c@chapter}{\newfloat{codelisting}{h}{lop}}{\newfloat{codelisting}{h}{lop}[chapter]}
\floatname{codelisting}{Listing}

\makeatother
\makeatletter
\@ifpackageloaded{caption}{}{\usepackage{caption}}
\@ifpackageloaded{subcaption}{}{\usepackage{subcaption}}
\makeatother

\ifLuaTeX
  \usepackage{selnolig}  
\fi
\usepackage[]{natbib}
\usepackage{bookmark}

\IfFileExists{xurl.sty}{\usepackage{xurl}}{} 
\hypersetup{
  pdftitle={Title},
  pdfauthor={Author 1; Author 2},
  pdfkeywords={3 to 6 keywords, that do not appear in the title},
  colorlinks=true,
  linkcolor={blue},
  filecolor={Maroon},
  citecolor={Blue},
  urlcolor={Blue},
  pdfcreator={LaTeX via pandoc}}

\newcommand{\anon}{1}

\newtheorem{Theorem}{Theorem}
\newtheorem{prop}{Proposition}
\newtheorem{lemma}{Lemma}
 
\newtheorem{example}{Example} 
\newtheorem{assumption}{Assumption} 

\newtheorem{remark}{Remark}

\newcommand{\argminD}{\arg\!\min~} 
\newcommand{\argmaxD}{\arg\!\max~}

\newcommand{\tmle}{\hat{\bm{\theta}}_{\lambda}}

\newcommand{\target}{{\bm \theta}_{H}}
\newcommand{\mle}{\hat{\bm{\theta}}}

\newcommand{\true}{{\bm \theta}_{0}}
\newcommand{\gauss}{{\bm \theta}_{g}}
\newcommand{\mse}{{\rm MSE}}

\newcommand{\tr}{{\rm trace}}

\newcommand{\sev}{\iota_{\text{min}}}

\begin{document}

\def\spacingset#1{\renewcommand{\baselinestretch}%
{#1}\small\normalsize} \spacingset{1}


\if1\anon
{
  \title{\bf Finite-Sample Risk Approximation and Risk-Consistent Tuning for Generalized Ridge Estimation in Nonlinear Models: Controlling Extreme Realizations
  }
  \author{$\text{Masamune Iwasawa}^{a}$\hspace{.2cm}\\
    $^{a}$ Doshisha University, 602-8580 Karasuma-higashi-iru, Imadegawa-dori,\\ Kamigyo-ku, Kyoto, Japan, Email: miwasawa@mail.doshisha.ac.jp}
  \maketitle
} \fi

\if0\anon
{
  \bigskip
  \bigskip
  \bigskip
  \begin{center}
    {\LARGE\bf Title}
\end{center}
  \medskip
} \fi

\bigskip
\begin{abstract} 
Maximum likelihood estimation in nonlinear models can exhibit substantial instability in finite samples when the data provide limited information about certain parameters. Such instability is driven by rare but extreme realizations of the estimator, which can dominate mean squared error (MSE) and lead to poor performance of conventional estimators. 
To address this issue, we consider ridge estimators that directly target MSE through regularization and thereby control extreme realizations. Developing this approach raises several challenges, including characterizing finite-sample MSE, selecting the penalty parameter, and achieving oracle risk performance.
We address these challenges using a unified framework based on a finite-sample approximation to the MSE.
Building on higher-order expansions, we derive an explicit first-order approximation to the finite-sample MSE of generalized ridge estimators in a broad class of nonlinear models.
This approximation reveals an explicit bias--variance trade-off and shows that generalized ridge estimators can improve upon the MLE in terms of MSE at the first-order level, even under target misspecification.
It also provides a tractable foundation for analyzing data-driven tuning, enabling us to show that the proposed MSE-based selection rule achieves oracle risk consistency.
Simulation results demonstrate that the proposed method substantially reduces the frequency and impact of extreme realizations, leading to large improvements in finite-sample risk relative to both the maximum likelihood estimator and cross-validation-based methods. 
An empirical illustration shows that the proposed MSE-based tuning approach can stabilize first-stage propensity score estimation and reveal sensitivity in subsequent treatment effect estimates that remains hidden under conventional estimators.

\end{abstract}

\noindent%
\vfill

\newpage
\spacingset{1.8} 

\section{Introduction}
Parameter estimation in nonlinear likelihood models can exhibit large mean squared error (MSE) in finite samples, even when the estimator is consistent and the sample size is moderate.
In many empirical applications, this arises when the data provide limited information about certain parameters.
Examples include duration models with heavy censoring \citep{andersen96}, Poisson models with excess zeros \citep{lambert92}, and discrete choice models with rarely chosen alternatives \citep{dejong19}.
Such situations occur in a wide range of empirical settings across economics, transportation, medicine, and political science.

A key feature of this phenomenon is that estimation error often manifests through rare but extreme realizations of the estimator.
In extreme cases such as separation in discrete choice models or monotone likelihood in duration analysis, the maximum likelihood estimator (MLE) may produce extremely large realizations in finite samples \citep{heinze01,heinze02}.
More generally, even in the absence of such extreme conditions, weak identification can lead to occasional but very large estimation errors.
Because MSE is based on squared deviations, even a small number of such extreme realizations can disproportionately affect finite-sample MSE.
This implies that finite-sample MSE can be driven not by typical estimation error but by rare extreme outcomes.
While previous studies have documented instability and large estimation errors, 
the role of such extreme outcomes in shaping finite-sample MSE has not been systematically analyzed in the context of nonlinear likelihood models.

These considerations suggest that controlling MSE directly is essential for improving finite-sample performance.
This motivates the use of regularization methods that explicitly target coefficient risk rather than predictive performance.
Selecting the penalty parameter to minimize MSE leads to stronger regularization in situations where extreme realizations would otherwise arise, thereby controlling large and unstable coefficient estimates that drive finite-sample behavior.

In this paper, we develop a framework that directly targets MSE as a measure of finite-sample performance using ridge-type estimators and data-driven tuning of the penalty parameter.
While ridge methods are commonly studied in high-dimensional settings, we focus on nonlinear likelihood models with moderate parameter dimension, where the MLE is well defined but finite-sample instability may still arise due to limited information in the data.

Developing this approach raises three main challenges.
First, because the estimator is defined implicitly in nonlinear likelihood models, its finite-sample MSE is difficult to characterize, making it unclear whether ridge regularization can improve MSE.
Second, even if such an expression were available, selecting the penalty parameter to directly target MSE would remain challenging, as the MSE depends on the unknown true parameter and cannot be evaluated from the data.
Third, it remains unclear whether, in this setting, the proposed data-driven selection method can attain performance comparable to that of the infeasible oracle choice.

We address these challenges using a unified approach based on a finite-sample approximation to the MSE, which serves as a common basis for both theoretical analysis and data-driven tuning.
Building on the higher-order expansion framework of \citet{rilstone96}, we derive an explicit first-order approximation to the MSE of generalized ridge estimators in a broad class of nonlinear models.

This approximation plays a key role in the analysis. 
It establishes that generalized ridge estimators can improve upon the MLE in terms of MSE at the first-order level, even under target misspecification, through an explicit bias--variance trade-off.
Moreover, it provides a tractable foundation for analyzing data-driven tuning rules.

Within this framework, we propose a data-driven method for selecting the penalty parameter that directly targets risk, defined as the trace of the MSE.
To this end, we approximate the generalized ridge estimator by a tractable shrinkage form, thereby enabling the application of Stein’s unbiased risk estimate (SURE).
We show that the resulting selector achieves oracle risk consistency.

The theoretical results also provide practical guidance for implementation.
In particular, the finite-sample MSE analysis informs the range of penalty parameters that are relevant for achieving MSE improvements and clarifies how the magnitude of the penalty interacts with the distance between the shrinkage target and the true parameter.

Simulation results illustrate the practical implications of the proposed approach.
In multinomial choice models with rare outcome categories, the proposed MSE-based selector substantially reduces extreme estimation outcomes relative to the MLE and cross-validation-based procedures, leading to large improvements in finite-sample performance.

We further illustrate the implications of the method through an empirical application examining the effects of maternal smoking intensity during pregnancy on birth weight.
In this application, some treatment categories contain relatively few observations, creating instability in first-stage propensity score estimation.
We show that the generalized ridge estimator with MSE-based tuning stabilizes propensity score estimation and can reveal sensitivity in subsequent treatment effect estimates that may remain hidden under conventional estimation methods.
Thus, the generalized ridge estimator can serve as a diagnostic tool for assessing the sensitivity of treatment effect estimates to such instability.

Overall, this paper contributes to the literature by developing a unified framework based on a finite-sample approximation to the MSE. 
Within this framework, we show that ridge regularization can improve MSE relative to the MLE, even under target misspecification, and that the proposed MSE-based selector achieves oracle risk consistency.

The remainder of the paper is organized as follows. Section~\ref{sec_literature} reviews related literature. Section~\ref{sec_estimator} introduces the estimator. Section~\ref{sec_finite} develops finite-sample approximations for MSE. Section~\ref{sec_lambda} presents the MSE-based penalty selection rule and establishes its theoretical properties. Section~\ref{sec_Implementation} discusses practical considerations for implementation. Section~\ref{sec_simulation} reports simulation results, Section~\ref{sec_application} presents the empirical illustration, and Section~\ref{sec_conclusion} concludes.

\section{Related Literature} \label{sec_literature}
Ridge-type shrinkage provides a broadly applicable approach to stabilizing estimation.
Although ridge is well known to reduce MSE in linear models \citep{theobald74}, its finite-sample MSE properties in nonlinear likelihood models remain less understood.
Existing contributions are typically model-specific.
For example, in binary logit models with a single categorical regressor, \citet{cessie92} and \citet{blagus20} show that ridge estimators can outperform the MLE in terms of estimation and prediction accuracy.
However, these analyses typically combine asymptotic arguments with parameter-wise comparisons and therefore do not provide a general characterization of finite-sample behavior.
This limitation is closely related to the difficulty of analyzing MSE in nonlinear likelihood models, where the estimator is defined implicitly and explicit expressions for MSE are generally unavailable.

A large literature studies data-driven selection of regularization parameters.
Cross-validation is widely used in practice to select regularization parameters based on predictive performance.
Information-criteria-based approaches, such as AIC and BIC, similarly rely on likelihood-based objectives.
While these methods are effective for prediction, they do not directly target coefficient risk measured by MSE.

An alternative approach is based on SURE \citep{stein81}, which provides an unbiased estimator of risk under Gaussian approximations.
SURE-based methods have been extensively studied in linear models and shrinkage estimation problems, and recent work has extended these ideas to more general settings, including regularized estimators \citep{abadie19}.
However, extending such approaches to nonlinear likelihood models is challenging because the estimator does not generally admit a tractable shrinkage representation, such as an explicit affine function of the data, making the direct application of SURE difficult.

Importantly, in nonlinear likelihood models, finite-sample MSE can be dominated by rare but extreme realizations of the estimator.
Standard selection methods, which focus on predictive performance, do not explicitly account for such tail-driven behavior.

Another strand of the literature studies higher-order bias correction for nonlinear estimators \citep{chen12,hahn24}. 
While such approaches reduce bias, they do not necessarily improve MSE because the associated increase in variance may dominate overall performance \citep{shao95}. 
This highlights the distinction between bias correction and directly targeting MSE as a measure of finite-sample performance, which is the focus of this paper.

A different line of work addresses settings in which the data contain excess zeros or rarely chosen alternatives, using models such as zero-inflated and related specifications \citep{lambert92}. 
While such approaches are effective in specific applications, they are inherently model-specific and do not address the broader issue of instability in likelihood-based estimation.

Methodologically, this paper relates to the higher-order expansion framework of \citet{rilstone96}, which provides stochastic expansions for nonlinear estimators.
Existing work has primarily focused on analyzing estimator properties, while applications to regularized estimators and data-driven tuning remain limited.
This paper contributes to this line of work by extending the framework to generalized ridge MLEs and applying it to the analysis of finite-sample MSE and data-driven penalty selection.

\section{Setup and Generalized Ridge Estimator}\label{sec_estimator}
Let $\{{\bm Z}_{i}\}_{i=1}^{N}$ be independent and identically distributed (i.i.d.)~observations with density $f({\bm Z}_{i}; {\bm \theta})$, where ${\bm \theta} \in \Theta \subset \mathbb{R}^p$. 
Depending on the application, $f$ may denote either an unconditional density of ${\bm Z}_{i}$ or a conditional density of the outcome given covariates. 
For notational simplicity, we use the notation $f({\bm Z}_{i}; {\bm \theta})$ in both cases. 
We assume that the true distribution of ${\bm Z}_i$ is given by $f({\bm Z}_{i}; {\bm \theta}_{0})$ for some ${\bm \theta}_{0} \in \Theta$. 
Define the log-likelihood as 
\[
		L_{N}({\bm \theta}) 
	= 
		N^{-1}\sum_{i=1}^{N} \log f({\bm Z}_{i}; {\bm \theta}).
\] 
The MLE $\mle$ maximizes $L_{N}({\bm \theta})$ over $\Theta$.

The use of ridge-type penalties in likelihood-based estimation has been studied in various contexts; see, for example, \citet{schaefer86} and \citet{cessie92}. 
In the linear regression literature, the term ''generalized ridge'' typically refers to estimators that allow for flexible penalty matrices (see, e.g., \citealp{hoerl70,hemmerle75}). 
Adopting the same terminology here, the generalized ridge MLE is
\begin{equation}
		\tmle 
	= 
		\underset{\bm{\theta} \in \Theta} {\argmaxD}
		\left\{ L_N(\bm{\theta}) - \lambda \|{\it \Lambda}(\bm{\theta} - \target)\|^2 \right\},
		\label{eq_TMLE}
\end{equation}
where $\lambda \geq 0$ is a tuning (penalty) parameter that controls the magnitude of regularization, 
${\it \Lambda}$ is a $p \times p$ multivalued weighting matrix such that ${\it \Lambda}'{\it \Lambda}$ is positive definite, and
$\target$ denotes a user-specified, non-random target value.

The penalty shrinks parameters toward the specified target $\target$, which may reflect prior knowledge. 
Later results show that the estimator improves on the  MLE in terms of finite-sample MSE, even when the target is misspecified.

This framework nests the standard ridge estimator as a special case when $\target = {\bm 0}$ and ${\it \Lambda}=I$. 
It also encompasses other ridge-type estimators, including fused ridge \citep{tibshirani05} and ridge-to-homogeneity estimators \citep{anatolyev20}. 
Moreover, it relates to targeted ridge procedures that shrink toward estimates from auxiliary data \citep{vanWieringen2022}, whereas our target values are fixed and non-random.

The MLE in nonlinear models may exhibit large finite-sample MSE even with substantial sample sizes, 
particularly when the data provide limited information about certain parameters. 
Two illustrative examples highlight such situations.

\begin{example}[Discrete choice models]\label{ex_logit}
In discrete choice models, the MLE can exhibit large finite-sample MSE when some categories are rarely chosen, 
leading to imprecise estimates and poor predictive performance \citep{ye14,dejong19}. 
This issue arises in a wide range of discrete choice models commonly used in empirical research.
\end{example}

\begin{example}[Proportional hazards models]\label{ex_duration}
In duration models with censoring, such as proportional hazards models \citep{cox72}, 
the MLE can exhibit substantial bias when survival probabilities are low 
and censoring is severe \citep{andersen96}.
\end{example}

These examples illustrate that large finite-sample MSE in nonlinear models is often driven by weak identification and limited information in the data, leading to nearly flat likelihood functions.
In such situations---such as cases close to separation or monotone likelihood---the estimator can take extremely large values, 
and these extreme realizations can dominate finite-sample MSE.

By shrinking estimates toward a target value, the generalized ridge MLE can stabilize estimation 
and reduce variance, thereby improving finite-sample MSE. 
These properties are formally established in the next section through a finite-sample approximation analysis.

 \section{Finite Sample Approximation}\label{sec_finite}
Unlike linear models, nonlinear likelihood estimators do not admit closed-form expressions, 
and their finite-sample MSE is generally intractable (see Online Appendix~G for detailed discussion).  
To address this, we develop a finite-sample approximation that yields tractable expressions 
for the MSE of the generalized ridge estimator.
We revisit finite-sample approximation approach of MSE for the MLE \citep{rilstone96}, then derive their counterparts for the generalized ridge MLE and compare the two.
All proofs are collected in Online Appendix C.

{\bf Notation.}
For notational simplicity, we write 
$f({\bm \theta}) = f({\bm Z}_i; {\bm \theta})$ 
whenever the dependence on the observation ${\bm Z}_i$ is clear from context.
Write the score as 
$
		{\bm S}({\bm \theta})
	=
		{\bm S}({\bm Z}_{i};  {\bm \theta})
	=
		\partial \log f({\bm Z}_i;{\bm \theta})/\partial{\bm \theta}
$
and
the Hessian as 
$
		H_1({\bm \theta})
	=
		H_1({\bm Z}_{i}; {\bm \theta})
	=
		\partial {\bm S}({\bm \theta})/\partial{\bm \theta}
$.
Let 
$
		H_{2}({\bm \theta})
	 = 
	 	H_2({\bm Z}_{i}; {\bm \theta})
$ 
denote the $p \times p^{2}$ matrix obtained by differentiating 
$H_{1}({\bm \theta})$ elementwise with respect to the $1 \times p$ vector ${\bm \theta}'$; 
that is, the $j$th element in the $l$th row of $H_{2}({\bm \theta})$ is the derivative of the 
$j$th element in the $l$th row of $H_{1}({\bm \theta})$ with respect to ${\bm \theta}'$. 
Similarly, define $H_{3}({\bm \theta})$ as the $p \times p^{3}$ matrix obtained by differentiating 
$H_{2}({\bm \theta})$ elementwise with respect to ${\bm \theta}'$.
Evaluate at ${\bm \theta}_0$ by writing ${\bm S}= {\bm S}({\bm \theta}_0)$, $H_j=H_j({\bm \theta}_0)$ for $j=1,2,3$. 
Define $Q=\{E(H_1)\}^{-1}$ and $Q_\lambda=\{E(H_1)-2\lambda{\it \Lambda}'{\it \Lambda}\}^{-1}$.
We measure MSE by the matrix $\mse(\tmle) = E\{(\tmle - \true)(\tmle - \true)'\}$.
Throughout the paper, ``MSE'' refers to this matrix-valued quantity, and risk is defined as its trace.
We also define the scalar risk as its trace:
$
R(\lambda) = \tr (\mse(\tmle)).
$
The Frobenius norm is $\|\cdot\|$.
Throughout, we use the notation $A>0$ to indicate that the symmetric matrix $A$ is positive definite.

Using these notations, the true value $\true$,  MLE $\mle$ and generalized ridge MLE $\tmle$ are assumed to satisfy:
\begin{equation*}
		E\{ {\bm S}({\bm \theta}_0)\}
	=
		{\bm 0},
	\qquad
		\frac{1}{N}\sum_{i=1}^N {\bm S}(\mle)
	=
		{\bm 0},
	\qquad
		\frac{1}{N}\sum_{i=1}^N {\bm S}(\hat{\bm \theta}_\lambda)
		-
		2\lambda{\it \Lambda}'{\it \Lambda}(\hat{\bm \theta}_\lambda-\target)
	=
		{\bm 0}.
\end{equation*}

We impose high-level rate and smoothness conditions (cf.~\citealp{rilstone96}):
\begin{assumption}\label{a_rate}
	$\tmle-{\bm \theta}_0 = O_p(N^{-1/2}) + O(\lambda)$.
\end{assumption}

\begin{assumption}\label{a_data}
	In a neighborhood of ${\bm \theta}_0$, $f({\bm z};{\bm \theta})$ is thrice continuously differentiable in ${\bm \theta}$; 
	$E({\bm S})=O(1)$, 
	$N^{-1}\sum_{i=1}^{N}{\bm S}=O_p(N^{-1/2})$; 
	$E(H_j)=O(1)$ and 
	$N^{-1}\sum_{i=1}^{N}H_j - E(H_j)=O_p(N^{-1/2})$ 
	for $j=1,2,3$.
\end{assumption}

\begin{assumption}\label{a_inverse}
	For some neighborhood of ${\bm \theta}_0$,
	$\big\{N^{-1}\sum_{i=1}^{N} H_1({\bm \theta})\big\}^{-1}=O_p(1)$ and
	$\big\{N^{-1}\sum_{i=1}^{N} H_1({\bm \theta})-2\lambda{\it \Lambda}'{\it \Lambda}\big\}^{-1}=O_p(1)$.
	Moreover $Q=O(1)$ and 
	$Q_\lambda=O(1)$.
\end{assumption}

\begin{assumption}\label{a_lipschitz}
	For $j=1,2,3$, 
	$\|H_j({\bm \theta})-H_j\|\le \|{\bm \theta}-{\bm \theta}_0\|\,M_i$ 
	in a neighborhood of ${\bm \theta}_0$ with
	 $E(|M_i|)\le C < \infty$.
\end{assumption}

The rate of convergence specified in Assumption~\ref{a_rate} is crucial 
for the stochastic expansions of bias and MSE, as it determines the order 
of certain terms appearing in the expansions; sufficient primitive conditions are given in Lemmas~A.1 and A.2 (Online Appendix A). 
Assumptions~\ref{a_data}--\ref{a_lipschitz} parallel \citet{rilstone96} with modifications for the ridge term (cf.~\citealp{yang15}). 
Assumption~\ref{a_inverse} is standard for finite-sample approximations; the parts involving $\lambda$ and $Q_\lambda$ are specific to the ridge problem.

\subsection{MSE of the Estimator}\label{subsec_MSE_est}
Let $\mse_{N^{-1}}(\cdot)$ denote the first-order approximation to the MSE (the expectation of the outer product of the leading term; Lemma~B.1). 
For the  MLE \citep[Prop.~3.4]{rilstone96},
\[
		\mse_{N^{-1}}(\mle)
	=
		N^{-1}Q\,E({\bm S}{\bm S}')\,Q.
\]
It coincides with the asymptotic variance and has no squared-bias term.

The following lemma establishes the first-order MSE of the generalized ridge MLE.

\begin{lemma}\label{theorem_MSE}
Under Assumptions~\ref{a_rate}--\ref{a_lipschitz},
\[
		\mse_{N^{-1}}(\tmle)
	=
		N^{-1}Q_\lambda E({\bm S}{\bm S}')Q_\lambda'
		+
		4\lambda^2 
		Q_\lambda {\it \Lambda}'{\it \Lambda}
		({\bm \theta}_0-\target)
		({\bm \theta}_0-\target)
		'{\it \Lambda}'{\it \Lambda} Q_\lambda'.
\]
\end{lemma}

The first term represents the variance (with $Q$ replaced by $Q_\lambda$) 
and is of order $O(N^{-1})$, while the second term corresponds to the squared 
first-order bias (cf.~Lemma~F.1 in Online Appendix F) and is of order $O(\lambda^{2})$.

Lemma~\ref{theorem_MSE} shows that the first-order MSE of $\tmle$ 
consists of two components: the variance term, which decreases with $\lambda$, 
and the squared bias term, which increases with $\lambda$.\footnote{
The finite-sample approximation is further supported by comparison with exact bias and MSE in linear models, 
where closed-form expressions are available (see Online Appendix~G).
} 
This variance--bias trade-off is absent for the MLE, and it is precisely what 
enables ridge-type estimators to improve upon the MLE in terms of MSE 
when $\lambda$ is chosen appropriately.

\begin{Theorem}\label{theorem_inadmissible}
Under Assumptions~\ref{a_rate}--\ref{a_lipschitz}:
\begin{enumerate}
	\item \label{theorem_inadmissible1}
	$\mse_{N^{-1}}(\tmle)=\mse_{N^{-1}}(\mle)$ when $\lambda=0$.
	
	\item \label{theorem_inadmissible4}
		If 
	$
		\int \sup_{{\bm \theta} \in {\mathscr N}}
		\|\partial f({\bm z}, {\bm \theta})/\partial {\bm \theta} \|
		d{\bm z} <\infty
	$
	and
	$
		\int \sup_{{\bm \theta} \in {\mathscr N}}
		\|\partial f({\bm z}, {\bm \theta})/\partial {\bm \theta} \partial {\bm \theta}' \|
		d{\bm z} <\infty
	$ 
	in a neighborhood of ${\bm \theta}_0$, 
	then for each $\true$ there exists $\bar\lambda>0$ such that for any $0<\lambda<\bar\lambda$,
$\mse_{N^{-1}}(\mle)-\mse_{N^{-1}}(\tmle)>0$.
\end{enumerate}
\end{Theorem}

The boundedness assumptions in Theorem~\ref{theorem_inadmissible}-\ref{theorem_inadmissible4}, 
together with the differentiability condition in Assumption~\ref{a_data}, 
ensure the information matrix equality (see Lemma 3.6 of \citealp{newey94}).

Theorem~\ref{theorem_inadmissible} demonstrates that, for sufficiently small $\lambda$, 
the generalized ridge MLE strictly improves the first-order MSE over the  MLE---regardless of target misspecification---across a broad class of nonlinear models.

Theorem~\ref{theorem_inadmissible} establishes pointwise dominance of 
the generalized ridge MLE over the  MLE at each $\true$, 
but this does not imply inadmissibility. 
A uniform bound valid for all $\true \in \Theta$ is not guaranteed 
without additional restrictions, such as boundedness of the parameter space.\footnote{
The possibility of establishing uniform dominance under a bounded parameter 
space does not contradict the classical admissibility result of 
\citet{james61}, which shows that the MLE is admissible when $p \leq 2$. 
Their result assumes an unbounded parameter space $\mathbb{R}^p$, 
whereas boundedness restricts the range of $\true$ and can alter 
admissibility properties.
In addition, they evaluate risk by the sum of MSEs across 
coordinates, whereas our results assess performance using the full MSE 
matrix for nonlinear likelihood-based estimators. }

Theorem~\ref{theorem_inadmissible} can be viewed as extending earlier results
for ridge estimators in binary logistic regression models
\citep{cessie92,blagus20} to a more general class of nonlinear models
with random covariates, using finite-sample approximations instead of
asymptotic analysis.

\begin{remark}\label{remark_theorem1}
Theorem~\ref{theorem_inadmissible} establishes the existence of an upper
bound $\bar{\lambda}$ such that the generalized ridge MLE improves on the
 MLE whenever $\lambda < \bar{\lambda}$. 
Although the theorem itself does not specify this bound, 
the proof shows that any $\lambda$ satisfying
\begin{equation}
	\lambda
	\leq
	\frac{\sev({\it \Lambda}' {\it \Lambda}) }
	{N  (\true- \target)' {\it \Lambda}' {\it \Lambda}{\it \Lambda}' {\it \Lambda} (\true- \target)}.  	
	\label{eq_upper_bound}
\end{equation}
is sufficient for the improvement result,
where
$\sev(A)$ denotes the smallest eigenvalue of $A$.
This clarifies that $\bar{\lambda}$ depends on the weighting matrix,
the distance between the true parameter and the target value, 
and the sample size.
The inequality also reveals that, with fixed $N$, 
the permissible magnitude of $\lambda$ increases 
as the target value $\target$ provides a better approximation to $\true$.
\end{remark}

When the target is correct (so that $\true = \target$), the right-hand in \eqref{eq_upper_bound}  becomes arbitrarily large.
Thus, the improvement of the MSE holds for any $\lambda > 0$ under the correct target.

The inequality in Remark~\ref{remark_theorem1} provides only a sufficient condition and does not yield an explicit value of $\bar{\lambda}$. 
The following proposition provides an illustrative characterization 
of $\bar{\lambda}$ in a special case.
This result is not required for the main analysis, 
but helps clarify how the admissible range of $\lambda$ depends 
on the accuracy of the MLE.

\begin{prop}\label{Prop_inadmissible}
Suppose the assumptions of Theorem~\ref{theorem_inadmissible} hold, and let
${\it \Lambda}=E(-H_1)^{1/2}$.
Then
$
	\tr(\mse_{N^{-1}}(\mle)-\mse_{N^{-1}}(\tmle))>0
$
holds 
\begin{enumerate}
\item  for any
$
		0
	<
		\lambda
	<
		\frac{E(\|Q \frac{1}{N}\sum_{i=1}^{N}S\|^{2})}
		{\|\true - \target\|^{2} - E(\|Q \frac{1}{N}\sum_{i=1}^{N}S\|^{2})}
$
if 
$
		\|\true-\target\|^{2}
	>
		E(\|Q \frac{1}{N}\sum_{i=1}^{N}S\|^{2})
$;

\item for all $\lambda>0$ if 
$
		\|\true-\target\|^{2}
	\le 
		E(\|Q \frac{1}{N}\sum_{i=1}^{N}S\|^{2})
$.
\end{enumerate}
\end{prop}

Proposition~\ref{Prop_inadmissible} provides an explicit expression for $\bar{\lambda}$ 
in the special case ${\it \Lambda} = E(-H_{1})^{1/2}$. 
Using the approximation
$
		-Q \frac{1}{N}\sum_{i=1}^{N}{\bm S} 
	= 
		- E(H_{1})^{-1} \frac{1}{N} \sum_{i=1}^{N}{\bm S} 
	\approx 
		\mle - \true,
$
the bound can be written approximately as
$
		\bar{\lambda} 
	\approx 
		\frac{E(\|\mle - \true\|^{2})}
		{\|\true - \target\|^{2} - E( \|\mle - \true\|^{2})}.
$

This representation clarifies the role of the MLE in determining the admissible range of $\lambda$. 
When the MLE is highly accurate (that is, $E(\|\mle-\true\|^{2})$ is small), the admissible range for $\lambda$ becomes correspondingly small. 
Conversely, when the MLE is imprecise, larger values of $\lambda$ can still yield improvements in MSE.

\subsection{MSE of Prediction}
Let $p({\bm \theta};{\bm z})$ denote a model-based prediction function, such as a density value, a regression mean, a choice probability, or a hazard.
Write $p({\bm \theta})=p({\bm \theta};{\bm z})$ and $p(\hat{\bm \theta})=p({\bm \theta})|_{{\bm \theta}=\hat{\bm \theta}}$.

We are interested in the MSE of the prediction $p(\hat{\bm \theta})$ relative to the true value $p(\true)$. 
By applying a second-order expansion around $\true$, the MSE of prediction admits an approximation similar to that of parameter estimation in Section~\ref{subsec_MSE_est}. 
This leads to the following result, which parallels Theorem~\ref{theorem_inadmissible} for parameter estimates.

\begin{Theorem} \label{theorem_inadmissible_prediction}
Suppose that the assumptions in Theorem \ref{theorem_inadmissible} hold. In addition, assume that 
$p({\bm \theta})$ is twice continuously differentiable with respect to ${\bm \theta}$ in a neighborhood ${\mathscr N}$ of $\true$, and that
	$
		\sup_{{\bm \theta} \in {\mathscr N}}
		\|\partial p({\bm \theta}) /\partial {\bm \theta} \|
		<
			\infty
	$
	and
	$
		\sup_{{\bm \theta} \in {\mathscr N}}
		\|\partial p({\bm \theta}) / \partial {\bm \theta} \partial {\bm \theta}' \|
		<
		\infty
	$
	for all $z \in {\mathcal Z}$, where ${\mathcal Z}$ is the range of $Z_{i}$.
Then, for each $\true$, there exists some value $\bar{\lambda}$ such that for any $0 < \lambda < \bar{\lambda}$, it holds, for all $z \in {\mathcal Z}$, that
\begin{align}
	\mse_{N^{-1}}\left(p(\mle)\right) 
	-	
	\mse_{N^{-1}}\left(p(\tmle)\right) > 0.	\notag
\end{align}
\end{Theorem}

Hence, for sufficiently small $\lambda$, the MSE improvement in parameter estimation carries over to prediction. 
The differentiability and boundedness requirements are mild and are met in the examples above.

 \section{Data-Driven choice of Penalty Parameters} \label{sec_lambda}

The previous section shows that the generalized ridge estimator 
can improve finite-sample MSE when the penalty parameter $\lambda$ is appropriately chosen. 
In practice, the penalty parameter $\lambda$ is typically selected by cross-validation, 
which optimizes predictive performance rather than MSE.
As a result, cross-validation methods may yield suboptimal penalty choices in settings where estimation is unstable. 
This distinction becomes important in nonlinear models, where finite-sample instability can arise.
To address this, we construct a data-driven selection rule that directly targets risk,
defined as the trace of the MSE, by approximating the risk function using a quadratic approximation to the likelihood combined with Stein’s identity.

We evaluate the penalty parameter based on the risk 
\( 
		R(\lambda)		
	=
		\tr(\mse(\tmle))		
\),
where the expectation is taken with respect to the sampling distribution,
treating $\lambda$ as fixed.

Our goal is to construct a data-driven selector that attains risk close to the infeasible optimal risk level
$
\inf_{\lambda \in \mathcal{L}_N} R(\lambda).
$
To achieve this, we construct a tractable approximation to $R(\lambda)$ and establish oracle risk consistency of the resulting data-driven selector.

Let ${\mathcal L}_{N} \subset [0, \bar\lambda_N]$ denote a closed and bounded set of candidate values. 
Remark~\ref{remark_theorem1} suggests that values of $\lambda$ leading to MSE improvement are of order $N^{-1}$. 
Motivated by this, we consider $\bar\lambda_N = O(N^{-1})$ and restrict attention to shrinking neighborhoods of zero. 
A theory-motivated choice of ${\mathcal L}_{N}$ is discussed in Subsection~\ref{sec_lambda_Guidance}.

 \subsection{MSE-based Selector} \label{sec_lambda_mse}
Assume
$\gauss \sim N(\true, V)$, where 
$
		V
	= 
		J^{-1}(\true)/N
$ 
and
$	
	J({\bm \theta})
	=
	-E\{H_{1}({\bm Z}_{i}; {\bm \theta})\}
$.
Using a local quadratic approximation, the generalized ridge estimator can be approximated by the shrinkage estimator
\begin{align*}
		\delta_{\lambda} (\gauss)
	&= 
		\underset{\bm{\theta} \in \Theta} {\argmaxD}
		\left\{ 
			L_{N}(\gauss) 
			- \tfrac{1}{2} ({\bm \theta} - \gauss)'  \hat{J} \: ({\bm \theta} - \gauss)  
			- \lambda \| {\it \Lambda} (\bm{\theta} - \target)\|^2 
		\right\} 
		\\
	&=
		 \target
		 +
		A_{\lambda}
		 (\gauss - \target),
\end{align*}
where
$
		\hat{J} = \hat{J}(\bar{\bm \theta})
	=
		-\frac{1}{N}\sum_{i=1}^{N} H_{1}({\bm Z}_{i};\bar{\bm \theta})
$ for some constant $\bar{\bm \theta}$ and
$A_{\lambda}= ( \hat{J} + 2 \lambda {\it \Lambda}'{\it \Lambda} )^{-1}	\hat{J}$.

Treating $\hat{J}$ as fixed, $\delta_\lambda(\gauss)$ is affine in $\gauss$.
Then, Stein's identity (\citealp{stein81}) implies
$
		E_{\gauss}\{		
			(\gauss - \true )'
			 \delta_{\lambda} (\gauss) 
		\}
	=
			\tr(V A_{\lambda})
$
and
$
		E_{\gauss}\{
			(\gauss - \true )'
			\gauss
		\}
	=
		\tr(V)
$,
where $E_{\gauss}$ denotes expectation under $\gauss \sim N(\true, V)$.
Then, the risk of $\delta_{\lambda} (\gauss)$ can be written as
\begin{align*}
		E_{\gauss}\{ \| \delta_{\lambda} (\gauss) - \true \|^{2}\}
	&=
		E_{\gauss}\{ \| \delta_{\lambda} (\gauss) - \gauss  \|^{2}\}
		+
		2\tr(V A_{\lambda})
		-
		\tr(V).
\end{align*}
This leads to the Stein’s unbiased risk estimate (SURE)
\[
		\text{SURE}_{\lambda}
	=
		\|\delta_{\lambda} (\gauss) - \gauss \|^{2} + 2 \tr(V A_{\lambda} ) - \tr(V).
\]

We approximate $\text{SURE}_{\lambda}$ by the plug-in estimator
\[
		\hat{R}(\lambda) 
	=
		\|\hat{\delta}_{\lambda} (\mle) - \mle \|^{2} + 2 \tr(\hat{V} \hat{A}_{\lambda} ) - \tr(\hat{V}),
\]
where
$
		\hat{\delta}_{\lambda} (\mle)
	=
		 \target
		 +
		\hat{A}_{\lambda}
		 (\mle - \target)	
$,
$
		\hat{A}_{\lambda} 
	=
		\{ \hat{J}(\mle) + 2 \lambda {\it \Lambda}'{\it \Lambda} \}^{-1}
		\hat{J}(\mle)	
$, 
and
$
		\hat{V}
	=
		 -\{\frac{1}{N}\sum_{i=1}^{N} H_{1}({\bm Z}_{i};\mle) \}^{-1}/N
$.

Then, the data-driven MSE-based selector $\hat{\lambda}$ is defined as
\[
		\hat{\lambda}
	\in
		\arg\min_{\lambda \in \mathcal{L}_N} \hat{R}(\lambda).
\]

\subsection{Risk Consistency of the MSE-based Selector}
To establish that the proposed selector achieves asymptotically optimal risk, we impose the following additional assumptions.
\begin{assumption}\label{a_mle_moment}
The MLE is $\sqrt{N}$-consistent:
$\mle-\true=O_p(N^{-1/2})$.
Moreover, there exists a constant $C<\infty$ such that
$
		\limsup_{N\to\infty} 
		E \| \sqrt{N}( \mle - \true) \|^{q} 
	\le 
		C,
$
for some integer $q$.
\end{assumption}
\begin{assumption}\label{a_regularityMLE}
The MLE admits the classical asymptotic linear expansion
with influence function $Q {\bm S}({\bm Z}_i;\true)$, i.e.,
$
		\sqrt{N}(\mle - \true)
	=
		-Q\frac{1}{\sqrt{N}}\sum_{i=1}^N {\bm S}({\bm Z}_i;\true)
		+
		o_p(1),
$
where $Q=\{E(H_1)\}^{-1}$.
The information matrix equality
$
		E({\bm S}{\bm S}')=-E(H_1)
$
holds
and
$
	E(\mathbf S\mathbf S')
$
is positive definite.
\end{assumption}
\begin{assumption}\label{as_score4thMoment}
$
	E\left\|
		H_1({\bm Z}_1;\true)
	\right\|^4
	<
	\infty
$
and
$
		E\|
			{\bm S}({\bm Z}_1;\true)
		\|^l
	<
		\infty
$  for some integer $l$.
\end{assumption}
\begin{assumption}\label{a_H1_moment}
Define
$\hat H_j({\bm \theta})=\frac{1}{N}\sum_{i=1}^N H_j({\bm Z}_i;{\bm \theta})$ for $j=1,2$.
There exists a constants $N_{0}$ and $C<\infty$
such that for all $N \geq N_{0}$,
$
		\sup_{{\bm \theta}\in  \Theta} 
		E\|\hat H_j({\bm \theta})\|^{k}
	 \le 
	 	C,
$
for $j=1,2$,
and
$
		\sup_{{\bm \theta} \in \Theta}\
		\sup_{\lambda\in [0,\bar\lambda_N]}
		E\|
			\{
				\hat H_1({\bm \theta})
				-
				2\lambda{\it \Lambda}'{\it \Lambda}
			\}^{-1}
		\|^{k} 
	\le 
		C,
$
for some integer $k$.
\end{assumption}

Assumption~\ref{a_regularityMLE} corresponds to standard regularity conditions 
ensuring asymptotic linearity and asymptotic normality of the MLE 
(see, e.g., \citealp{newey94}).
Assumption~\ref{a_mle_moment} imposes a moment condition on the scaled estimator, 
which allows occasional finite-sample instability but rules out heavy-tailed behavior strong enough to make the moments diverge. 
Such integrability conditions arise naturally when deriving uniform bounds on MSE.

The first condition in Assumption~\ref{a_H1_moment} ensures that the sample Hessian does not exhibit explosive heavy-tailed behavior.
The second condition allows occasional near-singularity of the sample Hessian, provided that $\lambda$ is not too small relative to such realizations.

Although $\text{SURE}_{\lambda}$ is an unbiased estimator of the risk (under the Gaussian approximation
$\gauss \sim N(\true,V)$ with $J$ treated as fixed),
the plug-in criterion $\hat R(\lambda)$ is no longer an unbiased estimator of the risk.
Nevertheless, we establish uniform convergence in probability:
\[
		\sup_{\lambda\in{\mathcal L}_{N}}
		\bigl|
			\hat R(\lambda) - R(\lambda)
		\bigr|
	\leq
		\sup_{\lambda\in{\mathcal L}_{N}}
		\bigl|
			\hat R(\lambda) - \tilde{R}(\lambda)
		\bigr|
		+
		\sup_{\lambda\in{\mathcal L}_{N}}
		\bigl|
			\tilde{R}(\lambda) - R(\lambda)
		\bigr|			
	\xrightarrow{p} 
		0,
\]
where
$ 
		\tilde R(\lambda)
	=
		\tr(
			\mse(\hat{\delta}_\lambda(\mle))
		)
$
and
the two terms on the right-hand side vanish by Lemmas B.7 and B.8 in Online Appendix B.

Combining this uniform convergence with the lower bound on the oracle risk established in Lemma B.12 in Online Appendix yields the following ratio consistency result.
\begin{Theorem}
\label{thm_ratio_consistency}
Assume Assumptions~\ref{a_rate}--\ref{a_H1_moment} 
with $q=16$, $l=8$, and $k=16$.
In addition, assume that Assumption~4 holds with $E(|M_i|^8)\le C<\infty$.
Let $\Theta$ be convex and denote 
$
	R(\hat\lambda)
	=
	R(\lambda)|_{\lambda=\hat\lambda}.
$
Then,
\[
	\frac{R(\hat\lambda)}
	{\inf_{\lambda\in{\mathcal L}_N}R(\lambda)}
	\xrightarrow{p}
	1.
\]
\end{Theorem}

Theorem~\ref{thm_ratio_consistency} shows that the proposed selector asymptotically attains the same risk level as the infeasible oracle choice of $\lambda$. 
Although the SURE formula is derived under a Gaussian approximation with $J$ treated as fixed, 
the asymptotic validity of the selector does not rely on these assumptions. 
These assumptions are used only as a device for constructing the criterion and do not restrict the data-generating process under which the result holds.

\section{Practical Guidance} \label{sec_Implementation}

The theoretical results developed in the previous sections provide guidance for the practical implementation of the generalized ridge MLE in empirical work. 
In particular, the finite-sample MSE analysis clarifies how key design choices affect estimator performance.

We focus on three key issues: 
(i) the theory-motivated determination of the search region for the penalty parameter, 
(ii) the interpretation and practical choice of the weighting matrix ${\it \Lambda}$, and 
(iii) the choice of the target parameter $\target$.

 \subsection{The Search Range of Penalty Parameter} \label{sec_lambda_Guidance}
To implement the minimization in practice,
we typically restrict the search set ${\mathcal L}_{N}$ to an interval,
$
	{\mathcal L}_{N}
	=
	[0,\lambda_{\max}],
$
and minimize $\hat{R}(\lambda)$ over a finite grid on ${\mathcal L}_{N}$.

A useful guideline for selecting $\lambda_{\max}$ is provided in Remark~\ref{remark_theorem1} that shows that the generalized ridge estimator improves on the MLE
whenever \eqref{eq_upper_bound} holds.
Although this upper bound depends on the unknown parameter $\true$ and is therefore infeasible,
it suggests that practically relevant values of $\lambda$ scale with $N^{-1}$
and depend on the magnitude of the deviation $\|\true-\target\|$.

Importantly, theoretical results do not identify the optimal value of $\lambda$ within the improvement region. 
Restricting the search set to a strict subset of this region may exclude values of $\lambda$ that yield larger MSE improvements within the admissible range. 
For this reason, we use \eqref{eq_upper_bound} to guide the construction of the search region, 
ensuring that it covers the improvement region without imposing overly restrictive bounds.

To obtain a feasible benchmark for $\lambda_{\max}$, we adopt a norm-based calibration.
Let $\delta=\true-\target$ and write $A={\it \Lambda}'{\it \Lambda}$.
Then,
$
	\delta'A^{2}\delta
	\geq
	\sev(A^{2})\|\delta\|^{2}
	=
	\sev(A)^{2}\|\delta\|^{2}
$.
Consequently, for any $r$ satisfying $\|\delta\| \geq r$, we have
$
	\frac{\sev(A)}
	{N\delta'A^{2}\delta}
	\leq
	\frac{1}{N\sev(A)r^{2}}.
$
Motivated by this inequality, we set
\begin{equation}
	\lambda_{\max}
	=
	\frac{1}{N\,\sev({\it \Lambda}'{\it \Lambda})\,r^{2}},
	\label{max_value_for_search}
\end{equation}
which provides a simple and feasible benchmark for the scale of the penalty parameter.
This choice yields a compact search region for grid-based minimization of $\hat{R}(\lambda)$,
while preserving the $N^{-1}$ scaling implied by the improvement condition in Remark~\ref{remark_theorem1}.

In practice, we fix $r$ at a prespecified constant.
In our numerical experiments in Section \ref{sec_simulation}, we set $r = 0.1$.  
The results are qualitatively unchanged under moderate variations in $r$ such as $r = 0.01$, which expands the admissible search range.
We also verified that using a larger value such as $r = 1$ yields similar qualitative conclusions.

 \subsection{The Choice of the Weighting Matrix ${\it \Lambda}$}

The role of ${\it \Lambda}$ is directly linked to the MSE approximation in Section \ref{sec_finite}, as it determines the directions in which variance reduction and bias are traded off.

Two weighting schemes are particularly natural in practice. 
The first is the Hessian-based weighting 
${\it \Lambda}=E(-H_1)^{1/2}$, 
which scales parameters according to the curvature of the likelihood. 
Because the Fisher information reflects the local sensitivity of the likelihood 
to parameter perturbations, this choice aligns shrinkage with directions 
in which the likelihood surface is weakly curved. 
In nonlinear models where instability often arises along such weakly 
identified directions, this weighting can improve the stabilizing effect 
of ridge regularization.

A second natural choice is the covariate-based weighting 
${\it \Lambda}=E(\bm X_i \bm X_i')^{1/2}$, 
which corresponds to scaling parameters according to the variability of 
the regressors. 
In linear regression models with zero-mean regressors, the expected Hessian 
is proportional to $E(\bm X_i \bm X_i')$, so the two weighting schemes coincide. 
In more general nonlinear models, however, the Hessian incorporates additional 
information about the curvature of the likelihood, which may provide a more 
informative scaling of the parameter space.

In practice, the population quantities appearing in these weighting matrices 
are replaced by sample analogues. 
In our simulations and empirical application, the Hessian-based weighting 
is implemented using the square root of the sample Hessian evaluated at the MLE, 
while the covariate-based weighting is constructed from the square root of the 
sample second-moment matrix of the regressors. 

One practical consideration is that, when the Hessian is weak or nearly singular, 
Hessian-based weighting leads to a larger value of $\lambda_{\max}$, 
which defines the upper bound of the search region and may in turn affect the range of the grid search.

 \subsection{Choice of the Target}

Another key element of the generalized ridge estimator is the choice of the shrinkage target $\target$. 
The results in Section~\ref{sec_finite} show that performance depends on the distance between $\target$ and $\true$, 
with closer targets yielding smaller bias and larger MSE improvements.

The target does not need to be correctly specified for ridge regularization to be beneficial. 
Even when $\target \neq \true$, the generalized ridge estimator can still achieve smaller MSE than the MLE 
when $\lambda$ is appropriately chosen. 
The data-driven selection of $\lambda$ developed in Section~\ref{sec_lambda_mse} helps control the resulting bias--variance trade-off.

In empirical applications, several practical choices for $\target$ are available. 
Common options include setting the target to zero, using values suggested by prior knowledge or theory, 
or constructing it from preliminary estimates based on simpler models or auxiliary data. 
For example, \citet{fessler19} discuss theory-based construction of shrinkage targets, 
while Section~\ref{sec_application} demonstrates construction based on auxiliary data.

 \section{Simulation} \label{sec_simulation}

This section evaluates the finite-sample performance of the generalized ridge MLE relative to the MLE, focusing on estimation and prediction MSE. 
All simulations are conducted in a multinomial logit model, where rarely chosen categories are known to induce large MSE for the MLE. 
According to Theorems~\ref{theorem_inadmissible} and \ref{theorem_inadmissible_prediction}, ridge regularization can mitigate this instability.

We consider a setting with three alternatives $Y_i \in \{1,2,3\}$ and $k=8$ covariates, with outcome probabilities 
$P(Y_i=1)=16/N$, $P(Y_i=2)=(N-16)/(2N)$, and $P(Y_i=3)=(N-16)/(2N)$.
Covariates are generated as ${\bm X}_i \sim N({\bm 0},\Sigma)$ with heterogeneous variances.
Full details are provided in Online Appendix~D.

This design departs from the standard asymptotic framework by allowing outcome probabilities to depend on $N$. 
It should therefore be interpreted as a finite-sample stress design, 
intended to isolate instability arising from rare outcomes rather than to approximate asymptotic behavior.
Finite-sample stability depends on $N$ through the conditioning of the realized Hessian, 
with smaller samples more likely to generate near-singular configurations and extreme estimation outcomes.

In the empirical illustration of Section~\ref{sec_application}, 
the smallest category contains  fewer observations relative to the number of parameters than in this simulation, 
indicating that the instability considered here is empirically relevant and may be even more pronounced in practice.

The generalized ridge estimator penalizes slope coefficients only. 
We consider moderate misspecification of the target, while results for correct specification and more severe misspecification are reported in Online Appendix~D.

Two weighting matrices are examined: an Hessian-based weighting using the sample Hessian, 
and a covariate-based weighting using $(N^{-1}\sum {\bm X}_i{\bm X}_i')^{1/2}$. 
We denote the corresponding estimators as $\text{GRIDGE}_H$ and $\text{GRIDGE}_X$.

The penalty parameter $\lambda$ is selected using either the proposed MSE-based method or likelihood-based $5$-fold cross-validation. 
The search region is ${\mathcal L}_N = (0,\lambda_{\max})$, with $\lambda_{\max}$ defined in \eqref{max_value_for_search} with $r=0.1$. 
Additional implementation details are reported in Online Appendix~D.
All results are based on 5{,}000 Monte Carlo replications.

\subsection{MSE of Estimates} \label{sec_simulation_mse}

\begin{table}[!h]
\centering
\caption{\label{Table_MSE_Cat_case2_lcv} Coefficient MSE}
\centering
\begin{threeparttable}
\begin{tabular}[t]{cccccc}
\toprule
\multicolumn{2}{c}{ } & \multicolumn{2}{c}{CV-based} & \multicolumn{2}{c}{MSE-based} \\
\cmidrule(l{3pt}r{3pt}){3-4} \cmidrule(l{3pt}r{3pt}){5-6}
SampleSize & MLE & $GRIDGE_H$ & $GRIDGE_X$ & $GRIDGE_H$ & $GRIDGE_X$\\
\midrule
100 & 1.0000 (127) & 0.9629 (122) & 0.8607 (111) & 0.2856 (29) & 0.3989 (42)\\
150 & 1.0000 (42) & 0.9046 (41) & 0.7608 (36) & 0.3009 (7) & 0.3748 (9)\\
200 & 1.0000 (17) & 0.7968 (14) & 0.7720 (13) & 0.3988 (6) & 0.4411 (6)\\
250 & 1.0000 (19) & 0.8805 (19) & 0.6921 (14) & 0.5044 (7) & 0.4525 (4)\\
300 & 1.0000 (5) & 0.9892 (5) & 0.9137 (3) & 0.8300 (2) & 0.8101 (1)\\
400 & 1.0000 (2) & 0.9932 (2) & 0.7739 (1) & 0.0306 (0) & 0.0792 (1)\\
500 & 1.0000 (4) & 1.0021 (4) & 0.7770 (2) & 0.7768 (2) & 0.8130 (2)\\
1000 & 1.0000 (1) & 0.9792 (1) & 0.9887 (1) & 0.1420 (0) & 0.2590 (1)\\
\bottomrule
\end{tabular}
\begin{tablenotes}[para]
\item \footnotesize Note: Each entry reports the trace of MSE for slope parameter estimates (excluding intercepts), normalized by the MLE value in each sample size. Numbers in parentheses are extreme counts, defined as the number of replications in which the maximum absolute coefficient estimation error exceeds a large threshold ($> 50$).
\end{tablenotes}
\end{threeparttable}
\end{table}

Table \ref{Table_MSE_Cat_case2_lcv} reports the risk, defined as the trace of the MSE of the slope parameter estimates (excluding intercepts), normalized by the corresponding MLE value.
Numbers in parentheses indicate the number of replications in which the maximum absolute coefficient error exceeds a large threshold ($>50$).\footnote{
Similar patterns arise for a wide range of thresholds (see Online Appendix D).
}
These events correspond to extreme estimation failures arising in nearly singular likelihood environments. 
Although infrequent, they can dominate average risk due to their large magnitude.

\textbf{Difference across Penalty Choice:}
Performance differences across selection methods are primarily driven by the control of extreme realizations. 
When such events occur, even infrequently, they can dominate average risk and outweigh typical estimation accuracy.

MSE-based selection substantially reduces the frequency of replications with extreme realizations relative to both the MLE and cross-validation.
In contrast, cross-validation improves typical performance but does not systematically eliminate extreme failures. 
As a result, MSE-based selection delivers markedly larger reductions in average risk.

The importance of controlling extreme realizations remains even in larger samples, as squared loss places disproportionate weight on large estimation errors.
Extreme events become less frequent as sample size increases, yet they occur irregularly across samples, so average risk need not decline smoothly.
Even when such realizations are rare, reducing their contribution to overall risk can substantially lower average risk.
By directly targeting risk, the proposed rule adapts the degree of regularization to the underlying risk structure.

To better understand the mechanism behind extreme realizations, Online Appendix D reports the condition numbers of the observed Hessian and the selected penalty parameters in replications with extreme realizations.
The results confirm that these failures arise in nearly singular likelihood environments and that MSE-based selection often tends to select relatively large penalty values in such settings.
This pattern reflects the need for stronger regularization to control extreme estimation behavior.
Importantly, the proposed method is not designed to select large penalties per se, but to minimize coefficient risk.
As a result, the selected penalty level depends on both the underlying risk structure and the penalty specification, and need not be large in general.

\textbf{Difference across Weighting Matrix:}
The relative performance of $\text{GRIDGE}_H$ and $\text{GRIDGE}_X$ depends not only on the weighting matrix itself but also on the selection rule used to determine the penalty parameter. 
The two weighting schemes regularize different directions in the parameter space: the Hessian-based (H-type) weighting is based on the observed curvature of the likelihood, whereas the covariate-based (X-type) weighting is based on the covariate second-moment structure.

Under MSE-based selection, the H-type weighting often performs particularly well. 
Because the MSE-based rule directly targets risk, it induces stronger effective regularization in directions where the observed curvature is weak. 
In this setting, curvature-aligned shrinkage effectively suppresses weakly identified directions and can eliminate extreme realizations.

In contrast, under CV-based selection, the X-type weighting frequently yields smaller risk than the H-type weighting. 
A plausible explanation is that the H-type penalty depends directly on the observed Hessian, which can be unstable in finite samples and particularly sensitive to near-singular curvature. 
Since cross-validation primarily optimizes typical likelihood performance rather than explicitly targeting risk, it may not select a sufficiently large penalty to offset this instability. 
As a result, weakly identified directions can persist, leading to occasional extreme realizations and higher average risk.

These findings indicate that the effectiveness of a given weighting matrix is closely tied to how the penalty parameter is chosen. 
Hessian-based weighting can be highly effective when combined with a risk-oriented selection rule, but may be more sensitive to finite-sample instability under likelihood-based cross-validation.

 \subsection{MSE of Prediction} \label{sec_simulation_prediction}

To assess predictive accuracy, we evaluate out-of-sample MSE using independently generated data $\{Y_{i}^{*}, {\bm X}_{i}^{*}\}$ from the same distribution as the estimation sample:
\[
	\frac{1}{N} \sum_{i=1}^{N} \sum_{j=1}^{J} 
	\{ P(Y_{i}^{*} = j \mid {\bm X}_{i}^{*}, \hat{\bm \beta}) - P(Y_{i}^{*} = j \mid {\bm X}_{i}^{*}, {\bm \beta}) \}^{2},
\]
where $\hat{\bm \beta}$ denotes the estimator of ${\bm \beta}$ (MLE or the generalized ridge MLE).

\begin{table}[!h]
\centering
\begin{threeparttable}
\centering
\caption{\label{table_predMSE_lcv_2} Prediction MSE}
\centering
\begin{tabular}[t]{cccccc}
\toprule
\multicolumn{2}{c}{ } & \multicolumn{2}{c}{CV-based} & \multicolumn{2}{c}{MSE-based} \\
\cmidrule(l{3pt}r{3pt}){3-4} \cmidrule(l{3pt}r{3pt}){5-6}
SampleSize & MLE & $\mathit{GRIDGE}_{H}$ & $\mathit{GRIDGE}_{X}$ & $\mathit{GRIDGE}_{H}$ & $\mathit{GRIDGE}_{X}$\\
\midrule
100 & 1.0000 & 0.8878 & 0.8661 & 1.0472 & 0.9773\\
150 & 1.0000 & 0.9017 & 0.9286 & 1.0350 & 0.9883\\
200 & 1.0000 & 0.9122 & 0.9552 & 1.0228 & 0.9944\\
250 & 1.0000 & 0.9241 & 0.9721 & 1.0383 & 0.9954\\
300 & 1.0000 & 0.9309 & 0.9751 & 1.0301 & 0.9973\\
400 & 1.0000 & 0.9446 & 0.9854 & 1.0146 & 0.9988\\
500 & 1.0000 & 0.9535 & 0.9942 & 1.0162 & 0.9995\\
1000 & 1.0000 & 0.9748 & 0.9974 & 1.0096 & 0.9999\\
\bottomrule
\end{tabular}
\begin{tablenotes}[para,flushleft]
\item \footnotesize Notes: Each entry reports the MSE of out-of-sample predictions, averaged over simulation replications. Values are normalized by the MLE prediction MSE (MLE = 1). 
\end{tablenotes}
\end{threeparttable}
\end{table}

\textbf{Overall predictive performance:}
Table \ref{table_predMSE_lcv_2} shows that likelihood-based cross-validation delivers the strongest predictive performance, particularly in smaller samples.

In contrast, MSE-based selection does not consistently improve average prediction accuracy and can lead to higher overall prediction loss. 
This reflects the fact that cross-validation directly optimizes predictive performance, whereas the MSE-based rule targets coefficient risk rather than prediction error.

\textbf{Tail prediction loss:}
Average performance measures do not fully capture behavior in replications where estimation becomes unstable. 
To investigate robustness against such extreme estimation outcomes, we additionally consider a tail-based performance measure.

For each configuration, replications are ranked according to the extremeness of the MLE estimates, measured by the maximum absolute value of the estimated coefficients (excluding intercepts), and prediction performance is evaluated by averaging out-of-sample loss over the top $(1-\alpha)$ fraction of replications.\footnote{
This measure is analogous to a conditional tail expectation (CVaR-type) criterion, focusing on predictive performance in replications with extreme realizations.
}

\begin{table}[!h]
\centering
\begin{threeparttable}
\centering
\caption{\label{table_tail_MSE2_lcv} Tail prediction loss}
\centering
\begin{tabular}[t]{cccccc}
\toprule
\multicolumn{2}{c}{ } & \multicolumn{2}{c}{CV-based} & \multicolumn{2}{c}{MSE-based} \\
\cmidrule(l{3pt}r{3pt}){3-4} \cmidrule(l{3pt}r{3pt}){5-6}
SampleSize & MLE & $\mathit{GRIDGE}_{H}$ & $\mathit{GRIDGE}_{X}$ & $\mathit{GRIDGE}_{H}$ & $\mathit{GRIDGE}_{X}$\\
\midrule
100 & 1.0000 & 0.9411 & 0.8584 & 0.2952 & 0.3979\\
150 & 1.0000 & 0.8857 & 0.7913 & 0.3466 & 0.3909\\
200 & 1.0000 & 0.7714 & 0.7579 & 0.5492 & 0.5771\\
250 & 1.0000 & 0.8899 & 0.7641 & 0.5163 & 0.4352\\
300 & 1.0000 & 0.8308 & 0.7975 & 0.8499 & 0.7383\\
400 & 1.0000 & 0.8507 & 0.8367 & 0.6072 & 0.6551\\
500 & 1.0000 & 0.9230 & 0.8089 & 0.8270 & 0.8340\\
1000 & 1.0000 & 0.8775 & 0.9494 & 0.8529 & 0.8910\\
\bottomrule
\end{tabular}
\begin{tablenotes}[para,flushleft]
\footnotesize
\item Notes: This table reports tail mean out-of-sample loss, computed over replications whose MLE out-of-sample loss exceeds the $\alpha$-quantile (here $\alpha=0.95$). All entries are normalized by the MLE loss.
\end{tablenotes}
\end{threeparttable}
\end{table}

The tail-based results differ markedly from average performance. 
MSE-based selection substantially reduces tail prediction loss relative to both the MLE and cross-validation, with gains concentrated in configurations where extreme realizations are more frequent, such as in smaller samples.

This reflects the distinct objectives of the selection rules. 
Cross-validation optimizes predictive likelihood in typical samples, whereas the MSE-based rule stabilizes coefficient estimates. 
By suppressing extreme coefficient realizations, the MSE-based rule reduces prediction errors in unstable replications.

Taken together, these results highlight a trade-off between average predictive accuracy and robustness to extreme estimation behavior. 
CV-based selection is preferable when overall predictive performance is the primary objective, whereas MSE-based selection can be advantageous when protection against rare but severe estimation failures is important.

 \section{Empirical Illustration} \label{sec_application}

To illustrate the practical implications of the proposed estimator, we revisit the maternal smoking data studied by \citet{cattaneo10}.
The treatment variable records the number of cigarettes smoked per day during pregnancy, categorized into six groups: $0$, $1$--$5$, $6$--$10$, $11$--$15$, $16$--$20$, and $21+$ cigarettes.
The outcome variable is birth weight.
Generalized propensity scores are estimated using a multinomial logit model with a series approximation for the covariates.

\begin{table}[!htbp]
\centering
\caption{Sample size by cigarette consumption}
\label{tab_cigarettes}
\begin{tabular}{lcccccc}
\toprule
& \multicolumn{6}{c}{Cigarettes per day}\\
& 0 & 1--5 & 6--10 & 11--15 & 16--20 & 21+ \\
\midrule
Sample size & 3652 & 198 & 326 & 61 & 221 & 42 \\ 
\bottomrule
\end{tabular}
\end{table}

Table~\ref{tab_cigarettes} reports the sample sizes across smoking categories.
The multinomial logit model contains $78$ parameters per treatment equation.
Given the sample sizes in Table~\ref{tab_cigarettes}, the number of observations per parameter is particularly small in the upper smoking categories, which contributes to near separation in the propensity score model.

The propensity model is estimated using both the MLE and the generalized ridge estimator with Hessian-based weighting and the MSE-based selector (hereafter $\text{GRIDGE}_{H, MSE}$).
The empirical illustration focuses on comparing these two estimators.

\begin{figure}[!t]
\begin{center}
 \includegraphics[width=140mm]{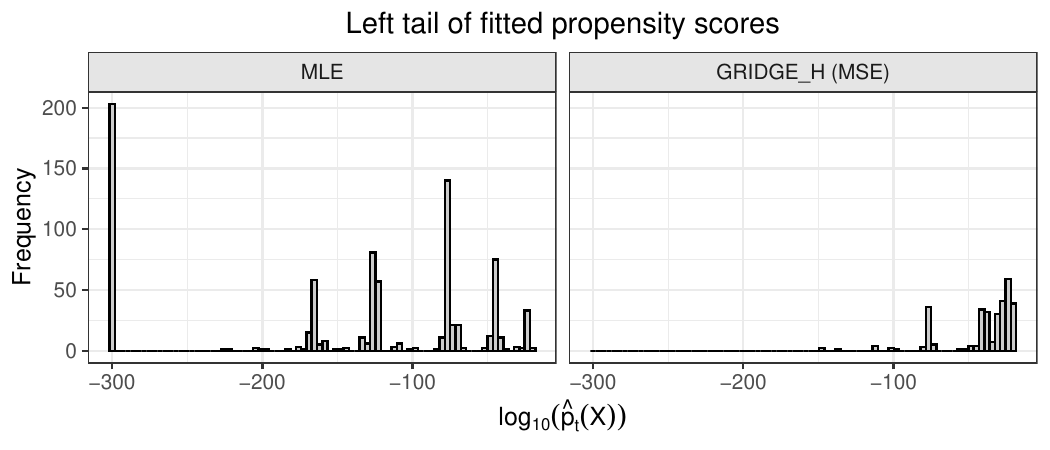}
\caption{\footnotesize
Left-tail distribution ($\log_{10}\hat p \le -20$) of fitted propensity scores under the multinomial logit MLE and the ridge estimator $\text{GRIDGE}_{H,MSE}$.
The ridge estimator substantially reduces the collapse of propensity scores toward zero.
}\label{fig_propensity}
\end{center}
\end{figure}

Figure~\ref{fig_propensity} illustrates instability in the propensity score estimation.\footnote{
For numerical stability, predicted propensity scores are truncated below at $10^{-8}$ when constructing inverse probability weights.
}
Under the MLE, near separation in the multinomial logit model leads to extremely large coefficient estimates and extremely small predicted probabilities.
In contrast, $\text{GRIDGE}_{H, MSE}$ stabilizes the estimation and mitigates the collapse of propensity scores toward zero.
Table~\ref{tab_diag_main} summarizes these diagnostic features.
The maximum absolute coefficient under the MLE exceeds $2.7\times10^{4}$, whereas the corresponding value under $\text{GRIDGE}_{H, MSE}$ is below $5\times10^{2}$.

\begin{table}[!htbp]
\centering
\caption{Diagnostic summary of propensity score instability}
\label{tab_diag_main}
\begin{tabular}{lcc}
\toprule
Statistic & MLE & $\text{GRIDGE}_{H, MSE}$ \\ 
\midrule
Max absolute coefficient & 27358.51 & 422.35 \\ 
  Number of floored propensities & 1042 & 913 \\ 
  Min Jacobian diagonal & $1.61\times10^{-5}$ & $1.78\times10^{-8}$ \\ 
  Selected $\lambda$ &  & 0.57224 \\ 
  Prediction error & 0.291 & 0.291 \\ 
\bottomrule
\end{tabular}
\end{table}

We next examine treatment effect estimation using inverse probability weighting estimators.
We consider two estimators: the marginal mean and the marginal $\tau$-quantile:
\[
	\hat{\mu}_{t}
	=
	\Bigg\{
		\sum_{i=1}^{n} 
		\frac{{\mathbbm 1}\{T_{i}=t\}}{\widehat{P}(T_{i}=t|{\bm X}_{i})}
	\Bigg\}^{-1}
	\sum_{i=1}^{n}
	\frac{{\mathbbm 1}\{T_{i}=t\} Y_{i}}{\widehat{P}(T_{i}=t| {\bm X}_{i})},
\] 
\[
	\hat{q}_{t}(\tau)
	=
	\underset{q} {\argminD} 
	\Bigg|
		\frac{1}{n}
		\sum_{i=1}^{n} 
		\frac{{\mathbbm 1} \{T_{i}=t\} (\mathbbm 1\{Y_{i} \leq q\} - \tau)}
		{\widehat{P}(T_{i}=t|{\bm X}_{i})}
	\Bigg|,
\]

These estimators depend critically on the estimated propensity scores.
While our theoretical analysis focuses on the finite-sample MSE properties of coefficient estimators, 
the empirical illustration examines how instability in the propensity score model propagates to treatment effect estimation through inverse probability weighting.

\begin{table}[!htbp]
\centering
\begin{threeparttable}
\caption{Treatment effect estimates for non-smokers and the heaviest smoking group}
\label{tab_TE_main}
\begin{tabular}{lcccc}
\toprule
   &\multicolumn{2}{c}{Mean treatment effect} &\multicolumn{2}{c}{0.1 quantile treatment effect} \\
   \cmidrule(lr){2-3} 
   \cmidrule(lr){4-5}
Group &  MLE & $\text{GRIDGE}_{H, MSE}$ &  MLE & $\text{GRIDGE}_{H, MSE}$ \\ 
\midrule
0 & 3402 & 3402 & 2734 & 2734 \\ 
   & (10) & (10) & (19) & (19) \\ 
21+ & 2978 & 2912 & 2084 & 1861 \\ 
   & (10) & (10) & (281) & (254527) \\ 
\bottomrule
\end{tabular}
\begin{tablenotes}[flushleft]
\footnotesize
\item[\hskip -\fontdimen 2 \font] Note: Numbers in parentheses are standard errors.
\end{tablenotes}
\end{threeparttable}
\end{table}

Table~\ref{tab_TE_main} reports estimates for the non-smoking group ($0$) and the heaviest smoking group ($21+$).
Mean effects are similar across estimators, whereas substantial differences arise in the lower tail ($0.1$ quantile).
In particular, for the $21+$ group the estimated Jacobian of the quantile estimating equation becomes extremely small under $\text{GRIDGE}_{H, MSE}$ (Table~\ref{tab_diag_main}), leading to a very large estimated standard error.

These results highlight an important distinction between first-step and second-step stability.
Instability in propensity score estimation can materially distort the precision of treatment effect estimates.
Depending on the realized sample, such distortion may appear either as deceptively small or excessively large standard errors, because extreme propensity scores can place disproportionate weight on a few observations, shifting the estimated quantile toward regions with very different outcome density.

In this sense, the ridge estimator does not necessarily provide uniformly more reliable inference than the MLE.
Rather, it serves as a robustness check by stabilizing the propensity score estimation and revealing how sensitive second-step inference is to instability in the first-step model.

Further implementation details and additional results are reported in Online Appendix~E. In this application, ridge estimators using cross-validation selectors produce estimates and standard errors similar to those obtained under the MLE, whereas the MSE-based selector reveals instability that is less apparent under those alternatives.

This difference reflects the distinct objectives of the selection methods. MSE-based tuning responds more directly to instability in coefficient estimation, particularly that arising from extreme realizations. We interpret this pattern as a diagnostic feature of the method in this application rather than as a general property.

 \section{Conclusion} \label{sec_conclusion}
This paper develops a unified framework based on a finite-sample approximation to the MSE for generalized ridge estimators in a broad class of nonlinear likelihood models. 

Using this framework, we show that generalized ridge estimators can improve upon the MLE in terms of MSE even under target misspecification. 
We also propose a data-driven selection rule based on a Stein-type approximation and establish that the resulting selector achieves oracle risk consistency within the same framework. 
Importantly, the validity of the selector does not rely on normality of the MLE, as the Gaussian approximation is used only as a device for constructing the criterion.

The simulation results highlight a key practical insight: finite-sample risk in nonlinear models is often dominated by rare but extreme realizations of the estimator. In such settings, average MSE is heavily influenced by extreme outcomes rather than typical estimation error. The proposed MSE-based selection rule effectively suppresses these extreme realizations and achieves substantial improvements in finite-sample risk, whereas cross-validation may fail to eliminate such instability because it targets predictive performance.

The empirical illustration further shows that stabilizing the propensity score model can reveal sensitivity in subsequent treatment effect estimation. This finding suggests that ridge regularization combined with the proposed MSE-based tuning serves as a diagnostic tool for detecting instability in inverse probability weighting and related multi-step estimators, particularly when such instability is associated with extreme realizations in propensity score estimation, rather than guaranteeing improved second-step inference.

Several directions remain for future research. 
First, while this paper focuses on generalized ridge estimators, it remains an open question whether other machine learning methods can effectively control tail-driven risk arising from extreme realizations. 
Second, although our analysis focuses on risk measured by MSE, alternative risk criteria may be more relevant in specific applications. 
Finally, extending the present framework to high-dimensional settings and studying the interaction between regularization and stability in such environments are important topics for future work.

\textbf{Acknowledgements:} 
We 
thank 
Kohtaro Hitomi, Sanghyeok Lee, Yoshihiko Nishiyama, Ryo Okui, Naoya Sueishi, Takahide Yanagi and 
the participants of the several meetings and conferences for their useful discussions.
We used ChatGPT 5, an AI language model, to assist with English proof-reading. Responsibility for the content remains entirely with the authors.

\textbf{Funding:} 
This work was supported by JSPS KAKENHI Grant Number 24K04826.

\textbf{Disclosure Statement:} 
The authors report there are no competing interests to declare.

\textbf{Data Availability Statement:}
The data used in the empirical illustration are publicly available from the replication materials of \citet{cattaneo10}. The data can be accessed via the authors’ GitHub repository at \url{https://github.com/mdcattaneo/replication-C_2010_JOE}.

\textbf{Declaration of generative AI and AI-assisted technologies in the manuscript preparation process:}
During the preparation of this work the author used ChatGPT 5, in order to assist with English proofreading, language polishing, and coding. After using this tool/service, the author reviewed and edited the content as needed and takes full responsibility for the content of the published article.

  \bibliography{tmle_2.bib}

@article{heinze01,
  title={A solution to the problem of monotone likelihood in Cox regression},
  author={Heinze, Georg and Schemper, Michael},
  journal={Biometrics},
  volume={57},
  number={1},
  pages={114--119},
  year={2001},
  publisher={Wiley Online Library}
}

@article{heinze02,
  title={A solution to the problem of separation in logistic regression},
  author={Heinze, Georg and Schemper, Michael},
  journal={Statistics in medicine},
  volume={21},
  number={16},
  pages={2409--2419},
  year={2002},
  publisher={Wiley Online Library}
}

@article{stein81,
  title={Estimation of the mean of a multivariate normal distribution},
  author={Stein, Charles M},
  journal={The Annals of Statistics},
  volume={9},
  number={6},  
  pages={1135--1151},
  year={1981},
  publisher={JSTOR}
}

@article{abadie19,
  title={Choosing among regularized estimators in empirical economics: The risk of machine learning},
  author={Abadie, Alberto and Kasy, Maximilian},
  journal={Review of Economics and Statistics},
  volume={101},
  number={5},
  pages={743--762},
  year={2019},
  publisher={MIT Press One Rogers Street, Cambridge, MA 02142-1209, USA journals-info~…}
}

@article{fessler19,
  title={How to use economic theory to improve estimators: Shrinking toward theoretical restrictions},
  author={Fessler, Pirmin and Kasy, Maximilian},
  journal={Review of Economics and Statistics},
  volume={101},
  number={4},
  pages={681--698},
  year={2019},
  publisher={MIT Press One Rogers Street, Cambridge, MA 02142-1209, USA journals-info~…}
}

@inproceedings{james61,
  title={Estimation with quadratic loss},
  author={James, William and Stein, Charles and others},
  booktitle={Proceedings of the fourth Berkeley symposium on mathematical statistics and probability},
  volume={1},
  pages={361--379},
  year={1961},
  organization={University of California Press}
}

@book{shao95,
  title={The jackknife and bootstrap},
  author={Shao, Jun and Tu, Dongsheng},
  year={1995},
  publisher={Springer Science$+$Business Media New York}
}

@article{cattaneo10,
  title={Efficient semiparametric estimation of multi-valued treatment effects under ignorability},
  author={Cattaneo, Matias D},
  journal={Journal of econometrics},
  volume={155},
  number={2},
  pages={138--154},
  year={2010},
  publisher={Elsevier}
}

@article{cox72,
  title={Regression models and life-tables},
  author={Cox, David R},
  journal={Journal of the Royal Statistical Society Series B: Statistical Methodology},
  volume={34},
  number={2},
  pages={187--202},
  year={1972},
  publisher={Wiley Online Library}
}

@article{andersen96,
  title={Estimating the survival function in the proportional hazards regression model: a study of the small sample size properties},
  author={Andersen, Per Kragh and Bentzon, Michael Weis and Klein, John P},
  journal={Scandinavian Journal of Statistics},
  volume={23},
  number={1},
  pages={1--12},
  year={1996},
  publisher={JSTOR}
}

@article{lambert92,
  title={Zero-inflated Poisson regression, with an application to defects in manufacturing},
  author={Lambert, Diane},
  journal={Technometrics},
  volume={34},
  number={1},
  pages={1--14},
  year={1992},
  publisher={Taylor \& Francis}
}

@article{ye14,
  title={Comparing three commonly used crash severity models on sample size requirements: Multinomial logit, ordered probit and mixed logit models},
  author={Ye, Fan and Lord, Dominique},
  journal={Analytic Methods in Accident Research},
  volume={1},
  pages={72--85},
  year={2014},
  publisher={Elsevier}
}

@article{newey94,
  title={Large sample estimation and hypothesis testing},
  author={Newey, Whitney K and McFadden, Daniel},
  journal={Handbook of Econometrics},
  volume={4},
  pages={2111--2245},
  year={1994},
  publisher={Elsevier}
}

@article{theobald74,
  title={Generalizations of mean square error applied to ridge regression},
  author={Theobald, Chris M},
  journal={Journal of the Royal Statistical Society Series B: Statistical Methodology},
  volume={36},
  number={1},
  pages={103--106},
  year={1974},
  publisher={Oxford University Press}
}

@article{hoerl70,
  title={Ridge regression: Biased estimation for nonorthogonal problems},
  author={Hoerl, Arthur E and Kennard, Robert W},
  journal={Technometrics},
  volume={12},
  number={1},
  pages={55--67},
  year={1970},
  publisher={Taylor \& Francis}
}

@article{hemmerle75,
  title={An explicit solution for generalized ridge regression},
  author={Hemmerle, William J},
  journal={Technometrics},
  volume={17},
  number={3},
  pages={309--314},
  year={1975},
  publisher={Taylor \& Francis}
}

@article{anatolyev20,
  title={A ridge to homogeneity for linear models},
  author={Anatolyev, Stanislav},
  journal={Journal of Statistical Computation and Simulation},
  volume={90},
  number={13},
  pages={2455--2472},
  year={2020},
  publisher={Taylor \& Francis}
}

@article{vanWieringen2022,
  title={Sequential learning of regression models by penalized estimation},
  author={van Wieringen, Wessel N and Binder, Harald},
  journal={Journal of Computational and Graphical Statistics},
  volume={31},
  number={3},
  pages={877--886},
  year={2022},
  publisher={Taylor \& Francis}
}

@article{schaefer86,
  title={Alternative estimators in logistic regression when the data are collinear},
  author={Schaefer, Robert L},
  journal={Journal of Statistical Computation and Simulation},
  volume={25},
  number={1-2},
  pages={75--91},
  year={1986},
  publisher={Taylor \& Francis}
}

@article{hahn24,
  title={Estimation of average treatment effects for massively unbalanced binary outcomes},
  author={Hahn, Jinyong and Liu, Xueyuan and Ridder, Geert},
  journal={Econometric Reviews},
  volume={43},
  number={6},
  pages={1--26},
  year={2024},
  publisher={Taylor \& Francis}
}

@article{yang15,
  title={A general method for third-order bias and variance corrections on a nonlinear estimator},
  author={Yang, Zhenlin},
  journal={Journal of Econometrics},
  volume={186},
  number={1},
  pages={178--200},
  year={2015},
  publisher={Elsevier}
}

@article{chen12,
  title={Finite-sample properties of the maximum likelihood estimator for the binary logit model with random covariates},
  author={Chen, Qian and Giles, David E},
  journal={Statistical Papers},
  volume={53},
  pages={409--426},
  year={2012},
  publisher={Springer}
}

@article{rilstone96,
  title={The second-order bias and mean squared error of nonlinear estimators},
  author={Rilstone, Paul and Srivastava, Virendra K and Ullah, Aman},
  journal={Journal of Econometrics},
  volume={75},
  number={2},
  pages={369--395},
  year={1996},
  publisher={Elsevier}
}

@article{blagus20,
  title={Mean squared error of ridge estimators in logistic regression},
  author={Blagus, Rok and Goeman, Jelle J},
  journal={Statistica Neerlandica},
  volume={74},
  number={2},
  pages={159--191},
  year={2020},
  publisher={Wiley Online Library}
}

@article{tibshirani05,
  title={Sparsity and smoothness via the fused lasso},
  author={Tibshirani, Robert and Saunders, Michael and Rosset, Saharon and Zhu, Ji and Knight, Keith},
  journal={Journal of the Royal Statistical Society Series B: Statistical Methodology},
  volume={67},
  number={1},
  pages={91--108},
  year={2005},
  publisher={Oxford University Press}
}

@article{cessie92,
  title={Ridge estimators in logistic regression},
  author={Le Cessie, S. and Van Houwelingen, J. C. },
  journal={Journal of the Royal Statistical Society Series C: Applied Statistics},
  volume={41},
  number={1},
  pages={191--201},
  year={1992},
  publisher={Oxford University Press}
}

@article{dejong19,
  title={Sample size considerations and predictive performance of multinomial logistic prediction models},
  author={de Jong, Valentijn MT and Eijkemans, Marinus JC and van Calster, Ben and Timmerman, Dirk and Moons, Karel GM and Steyerberg, Ewout W and van Smeden, Maarten},
  journal={Statistics in Medicine},
  volume={38},
  number={9},
  pages={1601--1619},
  year={2019},
  publisher={Wiley Online Library}
}

\end{document}